\documentclass[preprint,floatfix,showpacs,preprintnumbers,
amsmath,amssymb]{revtex4}
\usepackage{epsfig}

\usepackage{graphicx}
\usepackage{dcolumn}
\usepackage{bm}

\begin{document}


\title{Electron Spin Polarization in Resonant Interband Tunneling 
Devices}
\author{ A. G. Petukhov$^{1,2}$, D. O. Demchenko$^2$
\footnote{Present address: Physics Department, Georgetown University,
Washington, DC 20057},  and 
A. N. Chantis$^2$\footnote{Present address: Department of Chemical
and Materials Engineering, 
Arizona State University, Tempe, AZ 85287}}
\affiliation{$^1$Center for Computational Materials Science,
Naval Research Laboratory, Washington, DC 20375 \\
$^2$Physics Department, South Dakota School of Mines 
and Technology, Rapid City, SD 57701}

\date{\today}

\begin{abstract}
We study spin-dependent interband resonant tunneling in 
double-barrier InAs/AlSb/ Ga$_x$Mn$_{1-x}$Sb 
heterostructures. 
We demonstrate that  these structures
can be used as spin filters
utilizing spin-selective tunneling of electrons
through the light-hole resonant 
channel.
High 
densities of the spin polarized electrons injected into bulk 
InAs make spin resonant tunneling devices a viable alternative
for injecting spins into a semiconductor. 
Another striking feature of the proposed
devices is the possibility of inducing additional
resonant channels corresponding to the heavy holes. This
can be implemented by saturating the in-plane magnetization
in the quantum well.

\end{abstract}


\maketitle

One of the goals and challenges of the modern spintronics is the 
ability to create
stable sources of spin polarized electrons that can be injected
into the bulk of a semiconductor. One way of achieving this goal is
to inject spins from a ferromagnetic metal into a semiconductor
through a Schottky barrier 
\cite{rashba2000a,zhu2001a,hanbicki2002a}. Another alternative
consists in using all-semiconductor spin filtering devices.
One class of the proposed spin-filters utilizes the Rashba effect
in double-well resonant tunneling structures \cite{koga2002a}. 
Another class of these devices  uses  interband (or Zener)
spin-dependent tunneling in heterostructures comprising nonmagnetic
and magnetic semiconductors \cite{kohda2001a,johnston2002a}. 
The utilization of Zener tunneling 
in structures based on epitaxially grown III-V dilute magnetic semcionductors
(DMS)
is a necessary logical step in designing all-semiconductor
spin injection devices. Indeed, it has been proven experimentally that 
the electrons in III-V semiconductors
have remarkably long spin lifetimes while the holes tend to 
rapidly dissipate
their spin \cite{kikkawa1999a}.
Therefore, for futher spin manipulations, 
one needs the spin-polarized electrons 
rather than the holes, while  to date all known
III-V DMS are
$p$-type.   
The first spin-injection devices (Esaki diodes) based
on Zener tunneling of valence electrons from $p$-type ferromagnetic GaMnAs into
$n$-GaAs have
been already fabricated and successfully 
tested \cite{kohda2001a,johnston2002a}. 

\begin{figure}[htb]
\centerline{\epsfig{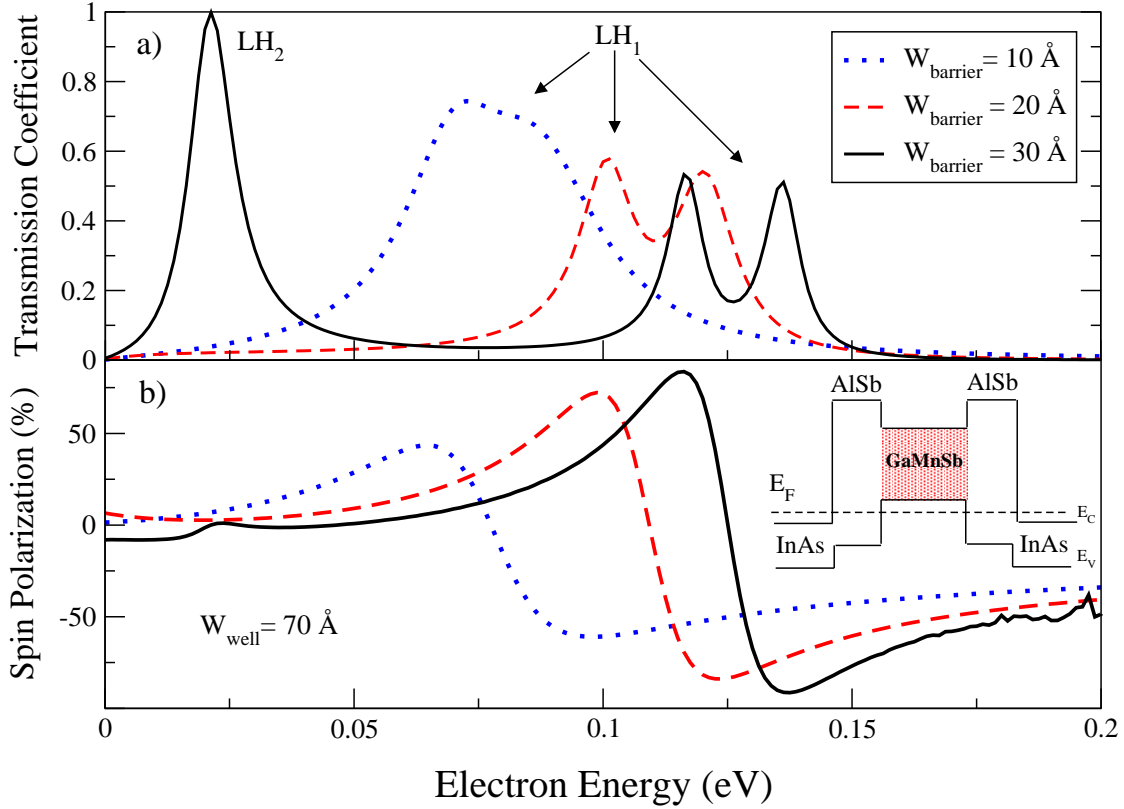}}
\caption{(a) Transmission coefficients of InAs/AlSb/GaMnSb 
heterostructure with  70~\AA~ quantum well
and various barrier widths; 
(b) Perpendicular single-electron spin polarizations at $\vec k_\parallel=0$}
\label{TransPolStr}
\end{figure}

In this Letter we 
consider theoretically another type of system that utilizes spin-dependent
{\em resonant} tunneling in magnetic heterostructures with type-II
broken-gap band alignment. These systems are
resonant interband tunneling devices (RITD) based on 
InAs/AlSb/GaMnSb/AlSb/InAs double-barrier hetorostructures
(DBH). A schematic band diagram of such a DBH is shown 
in the inset to Fig.\ref{TransPolStr}.
The band offset between InAs and GaMnSb leaves a 0.15 eV energy 
gap between 
the bottom of the conduction band in InAs and top of valence band in GaMnSb
\cite{liu1997a}. 
Therefore the electrons from  InAs emitter can tunnel 
through the hole states in the GaMnSb
quantum well into InAs collector. Since the quantized hole states in the
quantum well are spin-polarized the emerging electrons are expected
to be spin-polarized as well. 
Previous investigations of conventional (i. e. spin-independent)
interband resonant 
tunneling have been mainly focused on RITDs with 
GaSb quantum wells \cite {marquardt1996a,liu1997a} or
similar devices \cite{mendez1992a,mendez1992b,takamasu1992a}
and revealed quite robust operation in a wide temperature range.

The spin-filtering effect, or more precisely, the exchange splitting of the
light-hole channel has been observed experimentally in DBH with
semimetallic ErAs quantum wells \cite{brehmer1995a}.
The band diagram 
of the ErAs-based system is similar to that of the
proposed GaMnSb-based DBH
with the latter having an obvious 
advantage
of being ferro- rather than paramagnetic. 
Ga$_{1-x}$Mn$_x$Sb random alloys with Curie 
temperature $T_c\sim$ 25-30 K
have been grown and characterized \cite{matsukura2000a}.
Recently, much higher $T_c$ has been reported for Ga$_{1-x}$Mn$_x$Sb
digital alloys \cite{chen2002a}.
At the same time,  digital growth 
techniques are  proven to be very efficient for growing
high quality magnetic quantum wells \cite{crooker1995a}. This makes manufacturing
of GaMnSb-based spin-RITDs a technological reality. 

To describe spin-dependent interband resonant tunneling
in GaMnSb-based DBH we use standard $8\times 8$  ${\bf k\cdot p}$ Kane
Hamiltonians in the nonmagnetic InAs and AlSb regions \cite{talvar1994a} 
and a  generalized  
Kane Hamiltonian which accounts for magnetism in
Ga$_{1-x}$Mn$_x$Sb quantum well:
\begin{equation}
\label{Ham_tot}
H=H_e+H_h+H_{eh}.
\end{equation}
Here $H_{e}
=(\hbar^2k^2/2m^*)I$ is the electron Hamiltonian,
$I$ is $2\times 2$ unit matrix and we completely neglect
a small exchange splitting of the conduction band in Ga$_{1-x}$Mn$_x$Sb. 
The second term in Eq.(\ref{Ham_tot}) is the 
``Kohn-Luttinger+exchange'' hole Hamiltonian \cite{petukhov1996a,dietl2000a}:
\begin{equation}
H_h=
H_{KL}+\frac{3}{2}\Delta_{ex}(\vec\sigma\cdot\hat m),
\end{equation}
where $\vec\sigma\equiv(\sigma_x,\sigma_y,\sigma_z)$, $\sigma_\alpha$ 
are
Pauli matrices, $\hat{m}$ is the unit 
vector in the direction of magnetization,  $H_{KL}$
is a standard $6\times 6$ Kohn-Luttinger Hamiltonian for GaSb
\cite{talvar1994a}, and
$\Delta_{ex}$ is the exchange splitting of the light holes 
at $\vec k=0$. We will consider only saturation magnetizations
where $\Delta_{ex}=(5/6)\beta N_0 x$. Here 
$\beta$ is the $p-d$ exchange coupling constant, $N_0$ is the number
of cations per unit volume in Ga$_{1-x}$Mn$_x$Sb, and $x$ is Mn concentration.
The numerical value of $\Delta_{ex}\simeq $ 30 meV at $x=0.05$ 
is consistent with the Curie
temperature of bulk GaMnSb \cite{dietl2000a,matsukura2000a} $T_c\simeq$ 
25-30 K.
The third term in Eq.(\ref{Ham_tot}) describes  electron-hole 
coupling:
\begin{equation}
\label{Ham_eh}
H_{eh}^{\sigma}={P}\!\!\!\sum_{\alpha=x,y,z}{k_\alpha(|\alpha\sigma\rangle
\langle s\sigma|+ |s\sigma\rangle
\langle \alpha\sigma|
)},
\end{equation}
where $P$ is Kane's parameter related to the
matrix elements of the linear momentum operator
$(m/\hbar)P=\langle s|{p_x}|x\rangle
=\langle s|{p_y}|y\rangle=\langle s|{p_z}|z\rangle$,
and $|s\rangle$ and $|\alpha\rangle$ are $\vec k=0$ Bloch functions of the 
conduction and 
valence states respectively. We assume that $z$-axis is perpenicular
to the layers. 

To calculate the transmission coefficient we will use the transfer
matrix technique \cite{petukhov2002a} and represent the device as a 
stack of two-dimensional
flat-band interior layers with thickness $w_n=z_n-z_{n+1}$ and an average 
electrostatic potential $-eV_n$ starting
at $z=z_0=0$ and ending at $z=z_N$, where $z_N$ is the length
of the device.
The $0_{th}\equiv L$ and $(N+1)_{th}\equiv R$ layers are semi-infinite 
InAs emitter
($z<0$) and collector ($z>z_N$) having electrostatic potentials  $V_0=0$ 
and $-eV_{N+1}=-eV$ respectively, where $V$ is the bias applied to the
structure. We will take into account elastic processes only, i.e. assume
that the electron energy $E$ and lateral momentum $\vec k_\parallel$
are conserved. 
Substituting $k_z \rightarrow -i\hbar\partial/\partial z$ into
the Kane Hamiltonian one can solve
the Schr\"{o}dinger's equation in the $n_{th}$ flat-band region:
\begin{equation}
\label{psiMband}
\Psi_{n}(z)=
\sum_{\alpha=1}^{8}\left[
A_{n\alpha}^+ 
v(k_{n\alpha})
e^{ik_{n\alpha} z} + 
A_{n\alpha}^-
v(k_{n\alpha})
e^{-ik_{n\alpha} z}\right]
\end{equation}
This formula reflects the fact that the complex eigenvalues $\pm k_{n\alpha}$
always occur in pairs. The technique of finding $\pm k_{n\alpha}$
and eigenvectors $v(\pm k_{n\alpha})$ is described in \cite{petukhov2002a}.
 Using these quantities and the matching conditions
ensuring  wave function and 
current continuity across the device, we can construct the transfer matrix $M$
which relates the the wave-function amplitudes in the emitter and collector:
\begin{equation}
\label{M_part}
\left(\begin{array}{c} A_L^+\\A_L^-\end{array}\right)=
\left(\begin{array}{cc}
M_+ & M_{+-}\\
M_{-+}&M_{-}
\end{array}\right)
\left(\begin{array}{c} 
A_R^+\\0
\end{array}\right)
\end{equation}
The $16\times 16$ matrix $M$ in Eq. (\ref{M_part}) 
is partitioned in such a way
that $M_+$ expresses the amplitudes of the incident waves in
the emitter $A_L^+$ through those of the transmitted
waves $A_R^+$ in the collector.   The collector wavefunction is a 
superposition of these waves that are either
traveling solutions moving to the right (electron states) or
evanescent solutions decaying to the right (hole states).  

For any $k_{\parallel}$, $E$, and $-eV$, the $8\times 8$ transfer
matrix $M_+$ can be found and the transmission matrix 
can be calculated straightforwardly:
\begin{equation}
\label{trans_matrix}
t_{\sigma\sigma^\prime}=\left\{\begin{array}{cccccc}
\sqrt{j_{R\sigma^\prime}/
{j_{L\sigma}}}
\left(M_+\right)^{-1}_{\sigma^\prime\sigma}\!\!&\!\!,\!\!& 
\!\!\mbox{if}&\!\! j_{L\sigma}\!\! >0\!\! &\!\!\mbox{and} &\!\! j_{R\sigma^\prime}\!\! >0\\
0 &, & &  \mbox{otherwise} & & \end{array}\right .
\end{equation}
where $j_{L\sigma(R\sigma)}$ is the matrix element of the 
current operator for the electron with spin $\sigma$ in the
emitter (collector) \cite{petukhov2002a}. 
The transmission coefficient $T(\vec k_\parallel, E, eV)$ and
spin transmissivity $\vec{\cal S}(\vec k_\parallel, E, eV)$
\cite{slonczewski1989a}
can be calculated as well:
\begin{eqnarray}
T(\vec k_\parallel,E,eV)=\mathrm{Tr}(t\cdot t^\dagger)\\
\vec{\cal S}(\vec k_\parallel,E,eV)=
\mathrm{Tr}(t\cdot \vec \sigma \cdot t^\dagger)
\end{eqnarray}

The transmission coefficients for InAs/AlSb/GaMnSb 
DBH 
with 70~\AA-wide quantum well and 10, 20, and 30~\AA-wide 
barriers are shown in 
Fig. \ref{TransPolStr} (a).
The magnetization is saturated and directed along $z$-axis,
i.e. perpendicular to the 
layers. 
The spin splitting of the LH channels is well 
pronounced for the structure with wider
barriers (20 and 30~\AA). It is weakly
resolved as a shoulder  for the structure with 10~\AA~~ barriers. 
The barrier width, which is responsible for the width
of the spin-resolved peaks is therefore one of 
the critical 
parameters for the structures in question. 
This observation is well supported by the
calculated single-electron perpendicular spin polarizations ${\cal S}_z/T$
at $k_{\parallel}=0$,  shown in Fig. \ref{TransPolStr} (b)
for 70~\AA-wide quantum well and 
10, 20, and 30~\AA-wide barriers.
The maximum value of $p$ is 60\%
for the structure with 10~\AA barriers,
while 
for the structure with 30~\AA~barriers it reaches 95\%.

Similarly to the conventional InAs/GaSb RITDs the heavy hole (HH) 
resonant peaks are absent at $\vec k_{\parallel}=0$.
For perpendicular (or zero) magnetization
and a tunneling electron with $k_{\parallel}=0$, the 
$z$-component of the total angular momentum, $J_z$, 
is a good quantum number and, therefore, must be conserved.
Thus, tunneling of $s$-electrons near the condction
band minimum of InAs with $J_z={\pm}1/2$ through the hole 
states with $J_z={\pm}3/2$  is 
prohibited. 
Such a possibility exists for either finite 
$k_{\parallel}$ or
in-plane magnetization. As we will see below the former
is rather insignificant while the latter affects the interband
resonant tunneling in a drastic way.
For better insight we will treat our system by 
means of the tunneling Hamiltonian formalism \cite{gurvitz1993a} 
which gives an analytical expression for the transmission coefficient
at $\vec k_{\parallel}=0$. In this framework the system is described by two
coupled Schr\"{o}dinger equations:
\begin{eqnarray}
H^e_{\sigma}|\psi^e_{\sigma}\rangle + 
\sum_m{\hat V_{\sigma m}}|\varphi^h_m\rangle=
E|\psi^e_{\sigma}\rangle
\label{electron}\\
\sum_{m^\prime}H^h_{mm^\prime}|\varphi^h_{m^\prime}\rangle + \sum_{\sigma}
{\hat V_{m \sigma}}^{\dagger}|\psi^e_{\sigma}\rangle =  E|\varphi^h_m\rangle,
\label{holes}
\end{eqnarray}
Eq. (\ref{electron}) describes an electron with spin $\sigma=\pm1/2$
tunneling through the potential barrier (evanescent channel) which is
coupled with the confined hole states $\varphi_m^h$ by a mixing
potential $\hat{V}_{\sigma m}$. 
The basis of the localized hole states $|\varphi_m^h\rangle$ is defined
in terms of spherical harmonics with $m$ being the $z$-projection 
of the angular momentum onto the interface normal.
Thus matrix $H_{mm^\prime}$ is non-diagonal
for an arbitrary orientation of the magnetization with respect to the
interface. 
At $\vec k_{\parallel}=0$
the operator  $\hat{V}_{\sigma m}=-i\hbar P\delta_{\sigma,m}
\partial/\partial z$. Since $H^e_{\sigma}$ has continuous spectrum
the situation is typical for the appearance of Fano 
resonances \cite{fano1961a}.

We will
concentrate on the two particularly important cases
of magnetization perpendicular and parallel 
to the layers. 
Taking into account
only the first two  quantized hole levels 
$E_{1/2}$ (light hole) and $E_{3/2}$ (heavy hole) and
assuming that the barrier is
symmetric (i. e. $eV=0$), 
we  obtain the
the following expression for the transmission coefficient at $k_\parallel= 0$:
\begin{equation}
T_{\hat m}(E)=\frac{1}{2}T_0
\!\!\sum_{\sigma=\pm 1}\!\frac{
(\Sigma_{\hat m}^\sigma
(E) + \Delta_E)^2}{\left(\Sigma_{\hat m}^\sigma(E) + \Delta_E - \Gamma_E
\sqrt{R_0/T_0}\right)^2
+ \Gamma_E^2}
\label{transmission}
\end{equation}
where $T_0=|t_0|^2$ and $R_0=1-T_0$ are the ``bare'' transmission
and reflection coefficients, 
describing non-resonant and spin-independent electron
tunneling in the absence of the mixing potential, 
and the self-energy $\Sigma_{\hat m}^\sigma(E)$ is given by:
\begin{equation}
\Sigma_{\hat m}^\sigma(E)=\left\{\begin{array}{ll}E-E_{1/2}-
\frac{\textstyle 1}{\textstyle2}\sigma
\Delta_{ex},
\mbox{ ${\hat m} \parallel z$ } \\
E-E_{1/2}-\sigma\Delta_{ex}-
\frac{\textstyle 3\Delta_{ex}^2}{\textstyle 4(E-E_{3/2})},
\mbox{ ${\hat m} \parallel x$ }
\end{array}
\right.
\label{selfenergy}
\end{equation}
Here we  have introduced the inverse  elastic life-time of the
light-hole state $\Gamma_E\propto {P}^2$
and its energy shift due to the mixing potential 
$\Delta_E\propto {P}^2$. The expressions for $\Gamma_E$ and
$\Delta_E$ can be obtained straightforwardly, however they are rather
cumbersome and not important for our analysis. The only fact which
is important is that both $\Gamma_E$ and $\Delta_E$ are decreasing
functions of the barrier width.

\begin{figure}[htb]
\begin{center}
\mbox{\epsfig{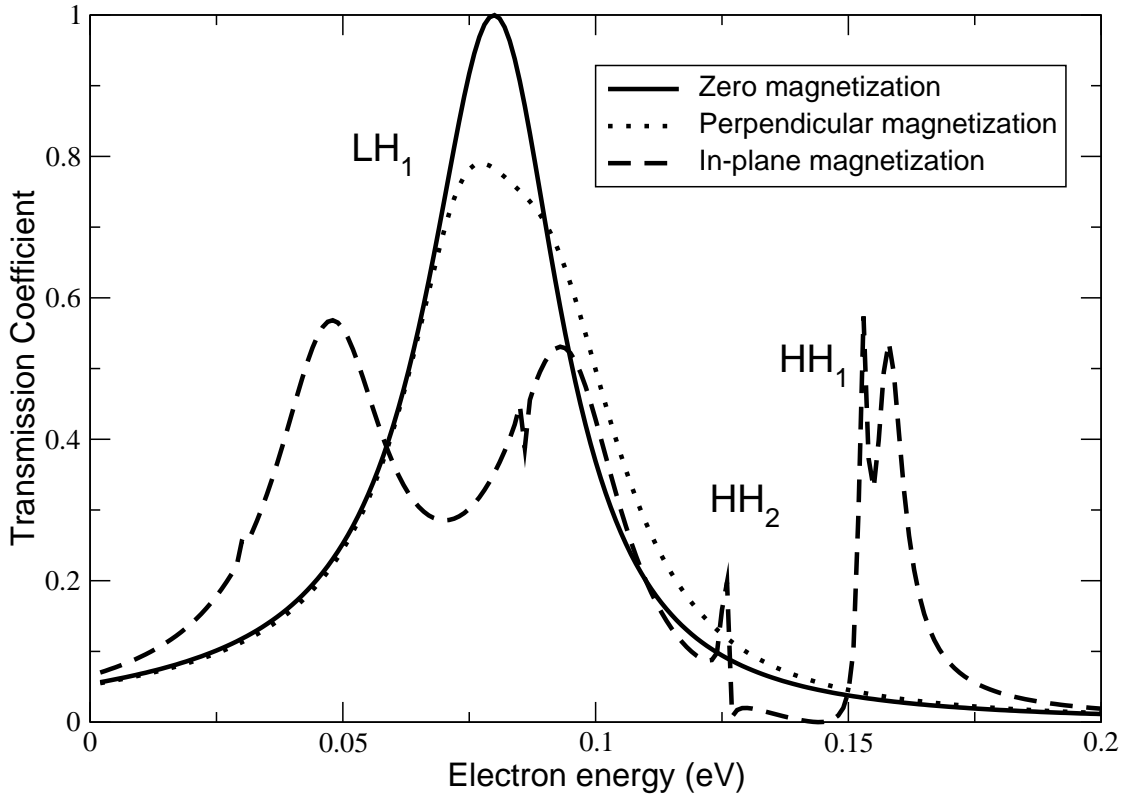}}
\end{center}
\caption{Transmission coefficients of InAs/AlSb/GaMnSb 
DBH with 70~\AA~quantum well and 10~\AA~barriers 
($\vec k_{\parallel}=0$).}
\label{transmission:fig}
\end{figure}

Eq. (\ref{selfenergy}) allows for rather meaningful and 
physically transparent interpretation
of the transmission coefficient, calculated numerically by means of the 
transfer matrix technique (Fig. \ref{transmission:fig}). 
First of all we note that Eq. (\ref{selfenergy}) 
describes a series
of Fano resonances and antiresonances
\cite{fano1961a,bowen1995a,klimeck2001a}. In the absence of the magnetization,
$\Sigma_+(E)=\Sigma_-(E)=\Sigma(E)$ and
we have only one resonance ($T(E)=1$) at 
$\Sigma + \Delta_E=\left(\Gamma_E/2\right)\sqrt{T/R}$ 
and anti-resonance ($T(E)=0$) at $\Sigma(E)+\Delta_E=0$, 
both corresponding to the light-hole channel. When the magnetization $\vec M$
is perpendicular to the layers, the light-hole (LH) resonance is exchange
split which leads to the perpendicular spin polarization of 
the transmitted electron
wave. Finally, the in-plane magnetization splits LH channel even
more strongly and induces another 
resonance-antiresonance pair corresponding to the heavy hole (HH) channel.
This channel, which is completely invisible for zero or perpendicular
magnetization, becomes very pronounced when $\vec M$ is in-plane.
This is a direct manifestation of the angular momentum selection rules
combined with the exchange enhancement of the effective $g$-factor
due to the localized Mn spins in the quantum well 
\cite{petukhov1996a,petukhov1998a}. 
 
\begin{figure}[htb]
\begin{center}
\mbox{\epsfig{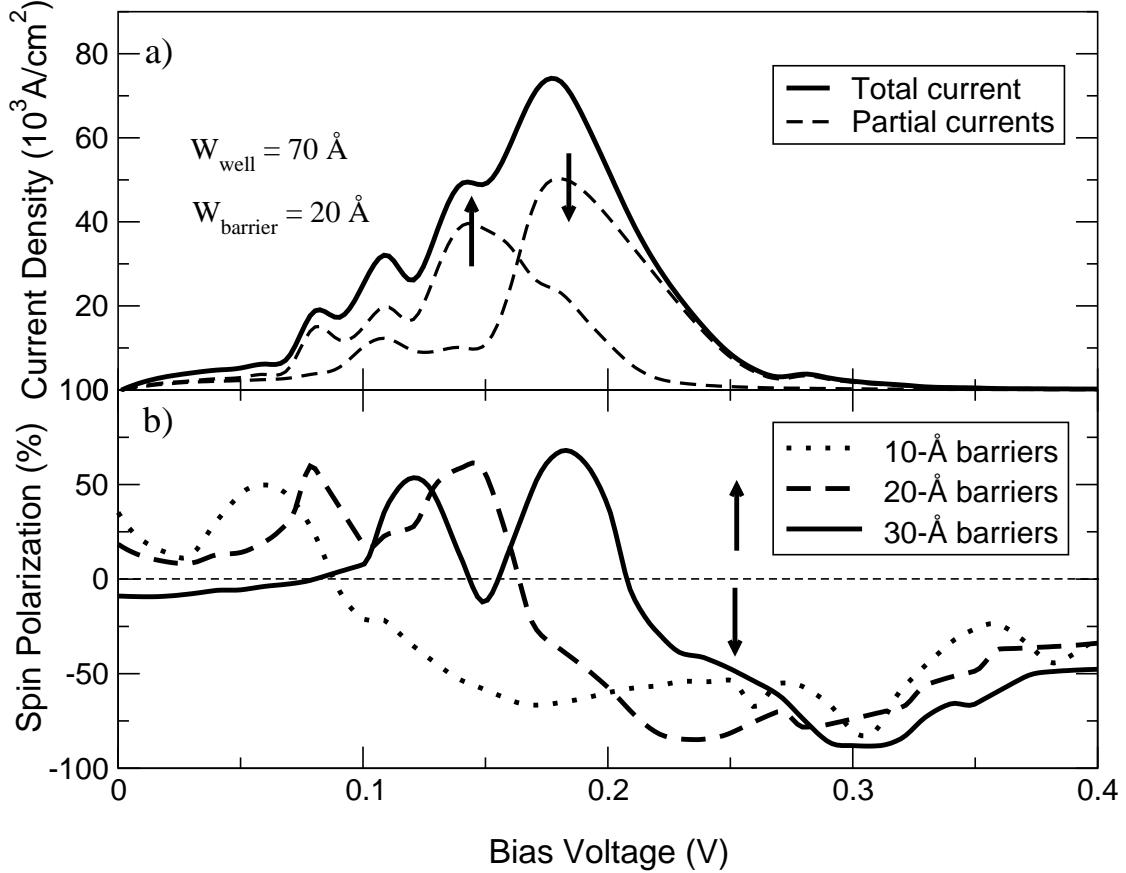}}
\end{center}
\caption{a) 
I-V 
curve for InAs/AlSb/GaMnSb DBH 
with  70~\AA~
quantum well, 20~\AA~barriers, and perpendicular magentization;
b) bias dependence of the 
current spin polarization $j_{s,z}/j$
for 70~\AA~quantum well, various barriers and perpendicular
magnetization.}
\label{I-V_pol}
\end{figure}

We now turn to a calculation charge $j$ and spin $\vec j_s$ current
densities \cite{duke1969a}:
\begin{eqnarray}
\label{current}
j\!\!=\!\!
\frac{e}{4 \pi^2 h}\!\!\int\!\!\!\!\int\!\! d\vec k_\parallel 
dE \left[f(E)\!-\!f(E\!+\!eV)\right]
T({\vec k}_{\parallel},E,eV)\\
\label{spin-current}
\vec j_s\!\!=\!\!
\frac{e}{4 \pi^2 h}\!\!\int\!\!\!\!\int\!\! d\vec k_\parallel 
dE \left[f(E)\!-\!f(E\!+\!eV)\right]
\vec{\cal S}({\vec k}_{\parallel},E,eV),
\end{eqnarray}
where $f(E)$ is the Fermi function.
Current-voltage characteristics calculated for $T$=4 K, 
20~\AA~barriers and a 70~\AA~quantum 
well are shown in 
Fig. \ref{I-V_pol} (a).
The current is calculated for perpendicular magnetization and the Fermi 
energy $E_F=0.04$ eV, which corresponds to electron 
concentration in InAs $n\sim 10^{18} cm^{-3}$, and
0.1 hole per Mn ion in the Ga$_{1-x}$Mn$_x$Sb quantum well at $x=0.05$.
The exchange splitting of the LH channel is resolved as a shoulder on the 
total current curve, which is a superposition of two very distinct
partial spin-up and spin-down currents.
This suggests that perpenicular
spin polarization $j_{s,z}/j$ of the 
tunneling current must be quite significant 
(Fig. \ref{I-V_pol} (b)). 
In this particular case (20~\AA~barriers)
$j_{s,z}/j$ reaches 90\%. The most 
striking result of our calculations is a sharp dependence of spin 
polarization on the
applied bias. As follows from Fig. \ref{I-V_pol} (b), 
one can drastically change both magnitude and sign of $j_{s,z}/j$
by applying external voltage.
Such controllable 
spin filtering is a remarkable feature of magnetic RITDs and may have 
significant potential 
for a variety of possible spin-injection applications \cite{prinz1990a}. 
Since the spin-split channels are better resolved for the structures 
with wider 
barriers (Fig. \ref{TransPolStr})
the spin polarization of the tunneling current is also higher for these structures. 
However, 
this effect is not as strong as we might expect, and the highest polarization 
values for different barrier widths are rather similar 
(Fig. \ref{I-V_pol} (b)). 

\begin{figure}[htb]
\begin{center}
\mbox{\epsfig{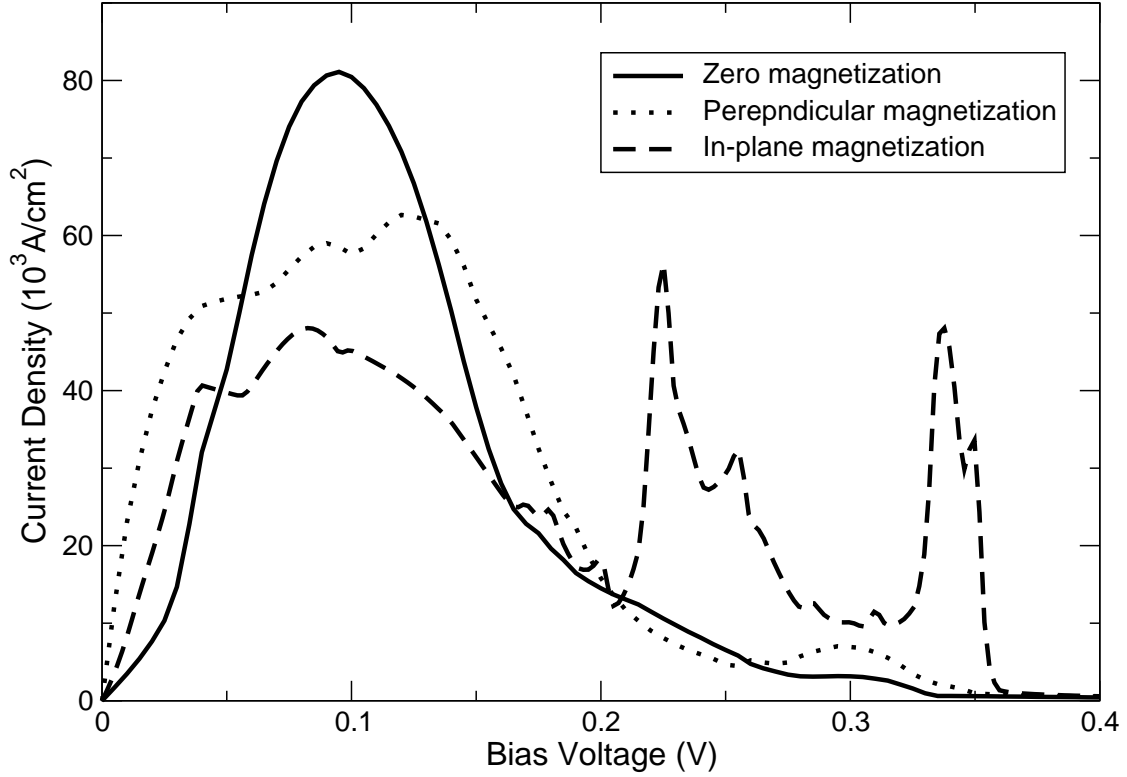}}
\end{center}
\caption{$I-V$ charactersitics of InAs/AlAs/GaMnSb DBH with
70~\AA~quantum well and
10~\AA~barriers,  for zero, perepndicular and in-palne 
magnetizations in the quantum well.
}
\label{I-V3}
\end{figure}

As we already mentioned, the in-plane magnetization affects resonant tunneling in
a dramatic way (Fig. \ref{transmission:fig}). 
Fig. \ref{I-V3} shows three current-voltage characteristics for zero, 
perpendicular, and parallel magnetizations, calculated for $T$ = 4 K nad 
the structure 
with 70~\AA~quantum well and 10~\AA~barriers.
The zero magnetization curve is similar to that of the
conventional RITDs and displays a very strong LH resonant channel
and a very weak feature stemming from the first HH state in the quantum well. 
The same is true for the  perpendicular
magnetization where the LH channel is  split into spin-up 
and spin-down subchannels but the HH peak remains very weak.
The most drastic changes of the $I-V$ characteristics occur for the
in-plane magnetization where, along with the splitting of the
light-hole channel, two new peaks related to the heavy hole states
in the quantum well emerge. This effect is due to the mixing of the
LH and HH channel at $\vec k_\parallel \simeq 0$ which results in the
lifting of the angular momentum selection rules \cite{petukhov1996a}.
The induced heavy-hole resonant channels have been clearly observed in
resonant tunneling through paramgnetic ErAs quantum wells in saturating in-plane
magnetic fields \cite{brehmer1995a}.
Even though nonzero $k_{\parallel}$ in the cases of perpendicular and 
zero magnetization 
also allows for tunneling of the 
electrons through the
heavy-hole states, the corresponding resonances are rather weak,
and are almost completely  washed out by the integration over $k_{\parallel}$
in Eq. (\ref{current}) (see also Ref. \cite{liu1997a}).

This work is supported by NSF Grant No 0071823 and by NRL under
ASEE-NAVY Summer Faculty Program.

\bibliographystyle{apsrev}
\bibliography{interband}

\begin{thebibliography}{28}
\expandafter\ifx\csname natexlab\endcsname\relax\def\natexlab#1{#1}\fi
\expandafter\ifx\csname bibnamefont\endcsname\relax
  \def\bibnamefont#1{#1}\fi
\expandafter\ifx\csname bibfnamefont\endcsname\relax
  \def\bibfnamefont#1{#1}\fi
\expandafter\ifx\csname citenamefont\endcsname\relax
  \def\citenamefont#1{#1}\fi
\expandafter\ifx\csname url\endcsname\relax
  \def\url#1{\texttt{#1}}\fi
\expandafter\ifx\csname urlprefix\endcsname\relax\def\urlprefix{URL }\fi
\providecommand{\bibinfo}[2]{#2}
\providecommand{\eprint}[2][]{\url{#2}}

\bibitem[{\citenamefont{Rashba}(2000)}]{rashba2000a}
\bibinfo{author}{\bibfnamefont{E.~I.} \bibnamefont{Rashba}},
  \bibinfo{journal}{Phys. Rev. B} \textbf{\bibinfo{volume}{62}},
  \bibinfo{pages}{R16267} (\bibinfo{year}{2000}).

\bibitem[{\citenamefont{Zhu et~al.}(2001)\citenamefont{Zhu, Ramsteiner,
  Kostial, Wassemeir, Schonherr, and Ploog}}]{zhu2001a}
\bibinfo{author}{\bibfnamefont{H.~J.} \bibnamefont{Zhu}},
  \bibinfo{author}{\bibfnamefont{M.}~\bibnamefont{Ramsteiner}},
  \bibinfo{author}{\bibfnamefont{H.}~\bibnamefont{Kostial}},
  \bibinfo{author}{\bibfnamefont{M.}~\bibnamefont{Wassemeir}},
  \bibinfo{author}{\bibfnamefont{H.-P.} \bibnamefont{Schonherr}},
  \bibnamefont{and} \bibinfo{author}{\bibfnamefont{K.~H.} \bibnamefont{Ploog}},
  \bibinfo{journal}{Phys. Rev. Lett.} \textbf{\bibinfo{volume}{87}},
  \bibinfo{pages}{016601} (\bibinfo{year}{2001}).

\bibitem[{\citenamefont{Hanbicki et~al.}(2002)\citenamefont{Hanbicki, Jonker,
  Itskos, Kioseoglou, and Petrou}}]{hanbicki2002a}
\bibinfo{author}{\bibfnamefont{A.~T.} \bibnamefont{Hanbicki}},
  \bibinfo{author}{\bibfnamefont{B.~T.} \bibnamefont{Jonker}},
  \bibinfo{author}{\bibfnamefont{G.}~\bibnamefont{Itskos}},
  \bibinfo{author}{\bibfnamefont{G.}~\bibnamefont{Kioseoglou}},
  \bibnamefont{and} \bibinfo{author}{\bibfnamefont{A.}~\bibnamefont{Petrou}},
  \bibinfo{journal}{Appl. Phys. Lett.} \textbf{\bibinfo{volume}{80}},
  \bibinfo{pages}{1240} (\bibinfo{year}{2002}).

\bibitem[{\citenamefont{Koga et~al.}(2002)\citenamefont{Koga, Nitta,
  Takayanagi, and Datta}}]{koga2002a}
\bibinfo{author}{\bibfnamefont{T.}~\bibnamefont{Koga}},
  \bibinfo{author}{\bibfnamefont{J.}~\bibnamefont{Nitta}},
  \bibinfo{author}{\bibfnamefont{H.}~\bibnamefont{Takayanagi}},
  \bibnamefont{and} \bibinfo{author}{\bibfnamefont{S.}~\bibnamefont{Datta}},
  \bibinfo{journal}{Phys. Rev. Lett.} \textbf{\bibinfo{volume}{88}},
  \bibinfo{pages}{126601} (\bibinfo{year}{2002}).

\bibitem[{\citenamefont{Kohda et~al.}(2001)\citenamefont{Kohda, Ohno, Takamura,
  Matsukura, and Ohno}}]{kohda2001a}
\bibinfo{author}{\bibfnamefont{M.}~\bibnamefont{Kohda}},
  \bibinfo{author}{\bibfnamefont{Y.}~\bibnamefont{Ohno}},
  \bibinfo{author}{\bibfnamefont{K.}~\bibnamefont{Takamura}},
  \bibinfo{author}{\bibfnamefont{F.}~\bibnamefont{Matsukura}},
  \bibnamefont{and} \bibinfo{author}{\bibfnamefont{H.}~\bibnamefont{Ohno}},
  \bibinfo{journal}{Jpn. J. Appl. Phys.} \textbf{\bibinfo{volume}{40}},
  \bibinfo{pages}{L1274} (\bibinfo{year}{2001}).

\bibitem[{\citenamefont{Johnston-Halperin
  et~al.}(2002)\citenamefont{Johnston-Halperin, Lofgreen, Kawakami, Young,
  Coldren, Gossard, and Awschalom}}]{johnston2002a}
\bibinfo{author}{\bibfnamefont{E.}~\bibnamefont{Johnston-Halperin}},
  \bibinfo{author}{\bibfnamefont{D.}~\bibnamefont{Lofgreen}},
  \bibinfo{author}{\bibfnamefont{R.~K.} \bibnamefont{Kawakami}},
  \bibinfo{author}{\bibfnamefont{D.~K.} \bibnamefont{Young}},
  \bibinfo{author}{\bibfnamefont{L.}~\bibnamefont{Coldren}},
  \bibinfo{author}{\bibfnamefont{A.~C.} \bibnamefont{Gossard}},
  \bibnamefont{and} \bibinfo{author}{\bibfnamefont{D.~D.}
  \bibnamefont{Awschalom}}, \bibinfo{journal}{Phys. Rev. B}
  \textbf{\bibinfo{volume}{65}}, \bibinfo{pages}{041306}
  (\bibinfo{year}{2002}).

\bibitem[{\citenamefont{Kikkawa and Awschalom}(1999)}]{kikkawa1999a}
\bibinfo{author}{\bibfnamefont{J.~M.} \bibnamefont{Kikkawa}} \bibnamefont{and}
  \bibinfo{author}{\bibfnamefont{D.~D.} \bibnamefont{Awschalom}},
  \bibinfo{journal}{Nature (London)} \textbf{\bibinfo{volume}{397}},
  \bibinfo{pages}{139} (\bibinfo{year}{1999}).

\bibitem[{\citenamefont{Liu et~al.}(1997)\citenamefont{Liu, Marquardt, Ting,
  and McGill}}]{liu1997a}
\bibinfo{author}{\bibfnamefont{Y.~X.} \bibnamefont{Liu}},
  \bibinfo{author}{\bibfnamefont{R.~R.} \bibnamefont{Marquardt}},
  \bibinfo{author}{\bibfnamefont{D.~Z.-Y.} \bibnamefont{Ting}},
  \bibnamefont{and} \bibinfo{author}{\bibfnamefont{T.~C.}
  \bibnamefont{McGill}}, \bibinfo{journal}{Phys. Rev. B}
  \textbf{\bibinfo{volume}{55}}, \bibinfo{pages}{7073} (\bibinfo{year}{1997}).

\bibitem[{\citenamefont{Marquardt et~al.}(1996)\citenamefont{Marquardt,
  Collins, Liu, Ting, and McGill}}]{marquardt1996a}
\bibinfo{author}{\bibfnamefont{R.~R.} \bibnamefont{Marquardt}},
  \bibinfo{author}{\bibfnamefont{D.~A.} \bibnamefont{Collins}},
  \bibinfo{author}{\bibfnamefont{Y.~X.} \bibnamefont{Liu}},
  \bibinfo{author}{\bibfnamefont{D.~Z.-Y.} \bibnamefont{Ting}},
  \bibnamefont{and} \bibinfo{author}{\bibfnamefont{T.~C.}
  \bibnamefont{McGill}}, \bibinfo{journal}{Phys. Rev. B}
  \textbf{\bibinfo{volume}{53}}, \bibinfo{pages}{13624} (\bibinfo{year}{1996}).

\bibitem[{\citenamefont{Mendez et~al.}(1992)\citenamefont{Mendez, Nocera, and
  Wang}}]{mendez1992a}
\bibinfo{author}{\bibfnamefont{E.~E.} \bibnamefont{Mendez}},
  \bibinfo{author}{\bibfnamefont{J.}~\bibnamefont{Nocera}}, \bibnamefont{and}
  \bibinfo{author}{\bibfnamefont{W.~I.} \bibnamefont{Wang}},
  \bibinfo{journal}{Phys. Rev. B} \textbf{\bibinfo{volume}{45}},
  \bibinfo{pages}{3910} (\bibinfo{year}{1992}).

\bibitem[{\citenamefont{Mendez}(1992)}]{mendez1992b}
\bibinfo{author}{\bibfnamefont{E.~E.} \bibnamefont{Mendez}},
  \bibinfo{journal}{Surf. Sci.} \textbf{\bibinfo{volume}{267}},
  \bibinfo{pages}{370} (\bibinfo{year}{1992}).

\bibitem[{\citenamefont{Takamasu{~et al.}}(1992)}]{takamasu1992a}
\bibinfo{author}{\bibfnamefont{T.}~\bibnamefont{Takamasu{~et al.}}},
  \bibinfo{journal}{Surf. Sci.} \textbf{\bibinfo{volume}{263}},
  \bibinfo{pages}{217} (\bibinfo{year}{1992}).

\bibitem[{\citenamefont{Brehmer et~al.}(1995)\citenamefont{Brehmer, Zhang,
  Schwarz, Chau, and Allen}}]{brehmer1995a}
\bibinfo{author}{\bibfnamefont{D.~E.} \bibnamefont{Brehmer}},
  \bibinfo{author}{\bibfnamefont{K.}~\bibnamefont{Zhang}},
  \bibinfo{author}{\bibfnamefont{C.~J.} \bibnamefont{Schwarz}},
  \bibinfo{author}{\bibfnamefont{S.~P.} \bibnamefont{Chau}}, \bibnamefont{and}
  \bibinfo{author}{\bibfnamefont{S.~J.} \bibnamefont{Allen}},
  \bibinfo{journal}{Appl. Phys. Lett.} \textbf{\bibinfo{volume}{67}},
  \bibinfo{pages}{1268} (\bibinfo{year}{1995}).

\bibitem[{\citenamefont{Matsukura et~al.}(2000)\citenamefont{Matsukura, Abe,
  and Ohno}}]{matsukura2000a}
\bibinfo{author}{\bibfnamefont{F.}~\bibnamefont{Matsukura}},
  \bibinfo{author}{\bibfnamefont{E.}~\bibnamefont{Abe}}, \bibnamefont{and}
  \bibinfo{author}{\bibfnamefont{H.}~\bibnamefont{Ohno}}, \bibinfo{journal}{J.
  Appl. Phys.} \textbf{\bibinfo{volume}{87}}, \bibinfo{pages}{6442}
  (\bibinfo{year}{2000}).

\bibitem[{\citenamefont{Chen et~al.}(2002)\citenamefont{Chen, Na, Wang, Luo,
  McCombe, Liu, Sasaki, Wojtowicz, Furdyna, Potashnik et~al.}}]{chen2002a}
\bibinfo{author}{\bibfnamefont{X.}~\bibnamefont{Chen}},
  \bibinfo{author}{\bibfnamefont{M.}~\bibnamefont{Na}},
  \bibinfo{author}{\bibfnamefont{S.}~\bibnamefont{Wang}},
  \bibinfo{author}{\bibfnamefont{H.}~\bibnamefont{Luo}},
  \bibinfo{author}{\bibfnamefont{B.~D.} \bibnamefont{McCombe}},
  \bibinfo{author}{\bibfnamefont{X.}~\bibnamefont{Liu}},
  \bibinfo{author}{\bibfnamefont{Y.}~\bibnamefont{Sasaki}},
  \bibinfo{author}{\bibfnamefont{T.}~\bibnamefont{Wojtowicz}},
  \bibinfo{author}{\bibfnamefont{J.~K.} \bibnamefont{Furdyna}},
  \bibinfo{author}{\bibfnamefont{S.~J.} \bibnamefont{Potashnik}},
  \bibnamefont{et~al.}, \bibinfo{journal}{Appl. Phys. Lett.}
  \textbf{\bibinfo{volume}{81}}, \bibinfo{pages}{511} (\bibinfo{year}{2002}).

\bibitem[{\citenamefont{Crooker et~al.}(1995)\citenamefont{Crooker, Tulchinsky,
  Levy, Awschalom, Garcia, and Samarth}}]{crooker1995a}
\bibinfo{author}{\bibfnamefont{S.~A.} \bibnamefont{Crooker}},
  \bibinfo{author}{\bibfnamefont{D.~A.} \bibnamefont{Tulchinsky}},
  \bibinfo{author}{\bibfnamefont{J.}~\bibnamefont{Levy}},
  \bibinfo{author}{\bibfnamefont{D.~D.} \bibnamefont{Awschalom}},
  \bibinfo{author}{\bibfnamefont{R.}~\bibnamefont{Garcia}}, \bibnamefont{and}
  \bibinfo{author}{\bibfnamefont{N.}~\bibnamefont{Samarth}},
  \bibinfo{journal}{Phys. Rev. Lett.} \textbf{\bibinfo{volume}{75}},
  \bibinfo{pages}{505} (\bibinfo{year}{1995}).

\bibitem[{\citenamefont{Talwar et~al.}(1994)\citenamefont{Talwar, Loehr, and
  Jogai}}]{talvar1994a}
\bibinfo{author}{\bibfnamefont{D.~N.} \bibnamefont{Talwar}},
  \bibinfo{author}{\bibfnamefont{J.~P.} \bibnamefont{Loehr}}, \bibnamefont{and}
  \bibinfo{author}{\bibfnamefont{B.}~\bibnamefont{Jogai}},
  \bibinfo{journal}{Phys. Rev. B} \textbf{\bibinfo{volume}{49}},
  \bibinfo{pages}{10345} (\bibinfo{year}{1994}).

\bibitem[{\citenamefont{Petukhov et~al.}(1996)\citenamefont{Petukhov,
  Lambrecht, and Segall}}]{petukhov1996a}
\bibinfo{author}{\bibfnamefont{A.~G.} \bibnamefont{Petukhov}},
  \bibinfo{author}{\bibfnamefont{W.~R.~L.} \bibnamefont{Lambrecht}},
  \bibnamefont{and} \bibinfo{author}{\bibfnamefont{B.}~\bibnamefont{Segall}},
  \bibinfo{journal}{Phys. Rev. B} \textbf{\bibinfo{volume}{53}},
  \bibinfo{pages}{3646} (\bibinfo{year}{1996}).

\bibitem[{\citenamefont{Dietl et~al.}(2000)\citenamefont{Dietl, Ohno,
  Matsukura, Cibert, and Ferrand}}]{dietl2000a}
\bibinfo{author}{\bibfnamefont{T.}~\bibnamefont{Dietl}},
  \bibinfo{author}{\bibfnamefont{H.}~\bibnamefont{Ohno}},
  \bibinfo{author}{\bibfnamefont{F.}~\bibnamefont{Matsukura}},
  \bibinfo{author}{\bibfnamefont{J.}~\bibnamefont{Cibert}}, \bibnamefont{and}
  \bibinfo{author}{\bibfnamefont{D.}~\bibnamefont{Ferrand}},
  \bibinfo{journal}{Science} \textbf{\bibinfo{volume}{287}},
  \bibinfo{pages}{1019} (\bibinfo{year}{2000}).

\bibitem[{\citenamefont{Petukhov et~al.}(2002)\citenamefont{Petukhov, Chantis,
  and Demchenko}}]{petukhov2002a}
\bibinfo{author}{\bibfnamefont{A.~G.} \bibnamefont{Petukhov}},
  \bibinfo{author}{\bibfnamefont{A.~N.} \bibnamefont{Chantis}},
  \bibnamefont{and} \bibinfo{author}{\bibfnamefont{D.~O.}
  \bibnamefont{Demchenko}}, \bibinfo{journal}{Phys. Rev. Lett.}
  \textbf{\bibinfo{volume}{89}}, \bibinfo{pages}{107205}
  (\bibinfo{year}{2002}).

\bibitem[{\citenamefont{Slonczewski}(1989)}]{slonczewski1989a}
\bibinfo{author}{\bibfnamefont{J.~C.} \bibnamefont{Slonczewski}},
  \bibinfo{journal}{Phys. Rev. B} \textbf{\bibinfo{volume}{39}},
  \bibinfo{pages}{6995} (\bibinfo{year}{1989}).

\bibitem[{\citenamefont{Gurvitz and Levinson}(1993)}]{gurvitz1993a}
\bibinfo{author}{\bibfnamefont{S.~A.} \bibnamefont{Gurvitz}} \bibnamefont{and}
  \bibinfo{author}{\bibfnamefont{Y.~B.} \bibnamefont{Levinson}},
  \bibinfo{journal}{Phys. Rev. B} \textbf{\bibinfo{volume}{47}},
  \bibinfo{pages}{10578} (\bibinfo{year}{1993}).

\bibitem[{\citenamefont{Fano}(1961)}]{fano1961a}
\bibinfo{author}{\bibfnamefont{U.}~\bibnamefont{Fano}}, \bibinfo{journal}{Phys.
  Rev.} \textbf{\bibinfo{volume}{124}}, \bibinfo{pages}{1866}
  (\bibinfo{year}{1961}).

\bibitem[{\citenamefont{Bowen et~al.}(1995)\citenamefont{Bowen, Frensley,
  Klimeck, and Lake}}]{bowen1995a}
\bibinfo{author}{\bibfnamefont{R.~C.} \bibnamefont{Bowen}},
  \bibinfo{author}{\bibfnamefont{W.~R.} \bibnamefont{Frensley}},
  \bibinfo{author}{\bibfnamefont{G.}~\bibnamefont{Klimeck}}, \bibnamefont{and}
  \bibinfo{author}{\bibfnamefont{R.~K.} \bibnamefont{Lake}},
  \bibinfo{journal}{Phys. Rev. B} \textbf{\bibinfo{volume}{52}},
  \bibinfo{pages}{2754} (\bibinfo{year}{1995}).

\bibitem[{\citenamefont{Klimeck et~al.}(2001)\citenamefont{Klimeck, Bowen, and
  Boykin}}]{klimeck2001a}
\bibinfo{author}{\bibfnamefont{G.}~\bibnamefont{Klimeck}},
  \bibinfo{author}{\bibfnamefont{R.~C.} \bibnamefont{Bowen}}, \bibnamefont{and}
  \bibinfo{author}{\bibfnamefont{T.~B.} \bibnamefont{Boykin}},
  \bibinfo{journal}{Superlattices and Microstructures}
  \textbf{\bibinfo{volume}{29}}, \bibinfo{pages}{187} (\bibinfo{year}{2001}).

\bibitem[{\citenamefont{Petukhov}(1998)}]{petukhov1998a}
\bibinfo{author}{\bibfnamefont{A.~G.} \bibnamefont{Petukhov}},
  \bibinfo{journal}{Appl. Surf. Sci.} \textbf{\bibinfo{volume}{123/124}},
  \bibinfo{pages}{385} (\bibinfo{year}{1998}).

\bibitem[{\citenamefont{Duke}(1969)}]{duke1969a}
\bibinfo{author}{\bibfnamefont{C.~B.} \bibnamefont{Duke}},
  \emph{\bibinfo{title}{Tunneling in Solids}} (\bibinfo{publisher}{Academic},
  \bibinfo{year}{1969}).

\bibitem[{\citenamefont{Prinz}(1990)}]{prinz1990a}
\bibinfo{author}{\bibfnamefont{G.~A.} \bibnamefont{Prinz}},
  \bibinfo{journal}{Science} \textbf{\bibinfo{volume}{250}},
  \bibinfo{pages}{1092} (\bibinfo{year}{1990}).

\end{thebibliography}

\end{document}